\documentstyle[sprocl]{article}
\def\Journal#1#2#3#4{{#1} {\bf #2}, #3 (#4)}

\def\PLB{{\em Phys.\ Lett.\ }B}

\def\ADNDT{{\em Atomic Data Nucl.\ Data Tables}}
\def\NPA{{\em Nucl.\ Phys.\ }A}
\def\ZPA{{\em Z. Phys.\ }A}

\def\be{\begin{equation}}
\def\ee{\end{equation}}
\def\bea{\begin{eqnarray}}
\def\eea{\end{eqnarray}}

\def\fig#1 #2 #3 #4 {\begin{figure} \centerline{\psfig{file=#2.ps,angle=-90,height=#3pt}} \caption[#1]{#4} \label{#1} \end{figure}}

\def\figt#1 #2 #3 #4 {\begin{figure}[t]
\centerline{\psfig{file=#2.ps,angle=-90,height=#3pt}} \caption[#1]{#4} \label{#1} \end{figure}}

\def\ess{\hskip.444444em plus .499997em minus .037036em}
\def\mss{\hskip.333333em plus .208331em minus .088889em}
\def\sen{\hbox{\scriptsize--}}
\def\eV{e\kern-.10emV }
\def\eVcm{e\kern-.10emV\kern-.15em,\mss}
\def\eVp{e\kern-.10emV\kern-.15em.\ess}
\def\eVpr{e\kern-.10emV) }

\input psfig

\begin{document}

\title{MASSES AND DEFORMATIONS OF NEUTRON-RICH NUCLEI}

\author{\underline{J. RAYFORD NIX} and PETER M\"OLLER}

\address{Theoretical Division, Los Alamos National Laboratory\\
Los Alamos, New Mexico 87545, USA\\
E-mail: nix@t2nix.lanl.gov\/ {\rm and\/} moller@moller.lanl.gov}


\maketitle

\abstracts{We have calculated the masses, deformations, and other properties of
8979 nuclei ranging from $^{16}$O to $^{339}$136 and extending from the proton
drip line to the neutron drip line on the basis of the 1992 version of the
finite-range droplet model.  The predicted quantities include the ground-state
mass, deformation, microscopic correction, odd-proton and odd-neutron spins and
parities, proton and neutron pairing gaps, binding energy, one- and two-neutron
separation energies, quantities related to $\beta$-delayed one- and two-neutron
emission probabilities, $\beta$-decay energy release and half-life with respect
to Gamow-Teller decay, one- and two-proton separation energies, and
$\alpha$-decay energy release and half-life.  For 1654 nuclei heavier than
$^{16}$O whose masses were known experimentally in 1989 and which were included
in the adjustment of model constants, the theoretical error is 0.669 M\eVp For
371 additional nuclei heavier than $^{16}$O whose masses have been measured
between 1989 and 1996 and which were not used in the adjustment of the model
constants, the theoretical error is 0.570 M\eVp}

\section{Introduction}\label{intro}

The accurate calculation of the ground-state mass and deformation of a nucleus
far from stability, such as one of the neutron-rich nuclei considered in this
conference, remains one of the most fundamental challenges of nuclear theory.
Toward this goal, two major approaches---which also allow the simultaneous
calculation of a wide variety of other nuclear properties---have been developed
(along with numerous semi-empirical formulas for masses alone).

At the most fundamental level, fully selfconsistent microscopic theories,
starting with an underlying nucleon-nucleon interaction, have seen progress in
both the nonrelativistic Hartree-Fock approximation and more recently the
relativistic mean-field approximation.  Although microscopic theories offer
great promise for the future, their current accuracies are typically a few
M\eVcm which is insufficient for most practical applications.  At the next level
of fundamentality, the macroscopic-microscopic method---where the smooth trends
are obtained from a macroscopic model and the local fluctuations from a
microscopic model---has been used in several recent global calculations that are
useful for a broad range of applications.

We will concentrate here on the 1992 version of the finite-range droplet
model,$\,$\cite{MNMS,MNK}\mss with particular emphasis on its reliability for
extrapolations to new regions of nuclei, but will also briefly discuss two other
models of this type.$\,$\cite{APDT,MS}\ess

\section{Finite-Range Droplet Model}\label{frdms} 

In the finite-range droplet model, which takes its name from the macroscopic
model that is used, the microscopic shell and pairing corrections are calculated
from a realistic, diffuse-surface, folded-Yukawa single-particle potential by
use of Strutinsky's method.$\,$\cite{S}\ess In 1992 we made a new adjustment of
the constants of an improved version of this model to 28 fission-barrier heights
and to 1654 nuclei with $N,Z \ge 8$ ranging from $^{16}$O to $^{263}$106 whose
masses were known experimentally in 1989.$\,$\cite{A}\ess The resulting
microscopic enhancement to binding for even-even nuclei throughout the periodic
system is shown in Fig.~\ref{enhab}.

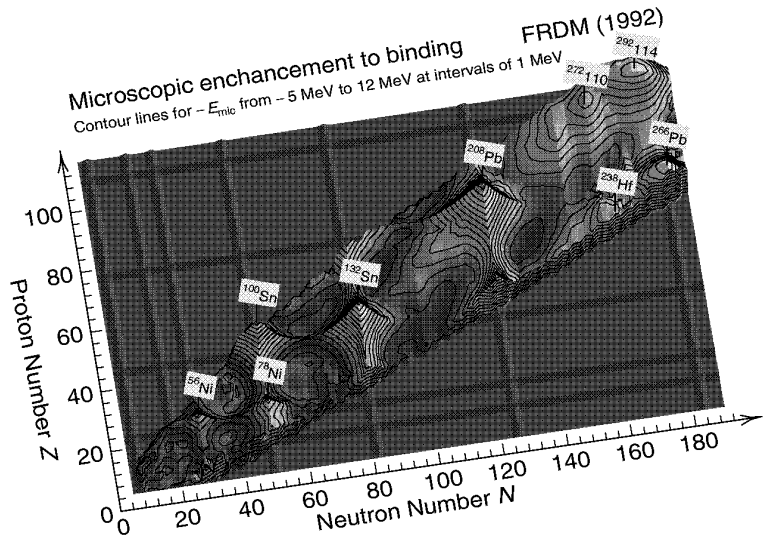
\fig enhab fig1 201 {Calculated additional binding energy of
even-even nuclei relative to the macroscopic energy of spherical nuclei,
illustrating the crucial role of microscopic corrections.}

This model has been used to calculate the ground-state mass, deformation,
microscopic correction, odd-proton and odd-neutron spins and parities, proton
and neutron pairing gaps, binding energy, one- and two-neutron separation
energies, quantities related to $\beta$-delayed one- and two-neutron emission
probabilities, $\beta$-decay energy release and half-life with respect to
Gamow-Teller decay, one- and two-proton separation energies, and $\alpha$-decay
energy release and half-life for 8979 nuclei with $N,Z \ge 8$ ranging from
$^{16}$O to $^{339}136$ and extending from the proton drip line to the neutron
drip line.$\,$\cite{MNMS,MNK}\ess These tabulated quantities are available
electronically on the World Wide Web at the Uniform Resource Locator {\tt
http://t2.lanl.gov/publications/publications.html}.

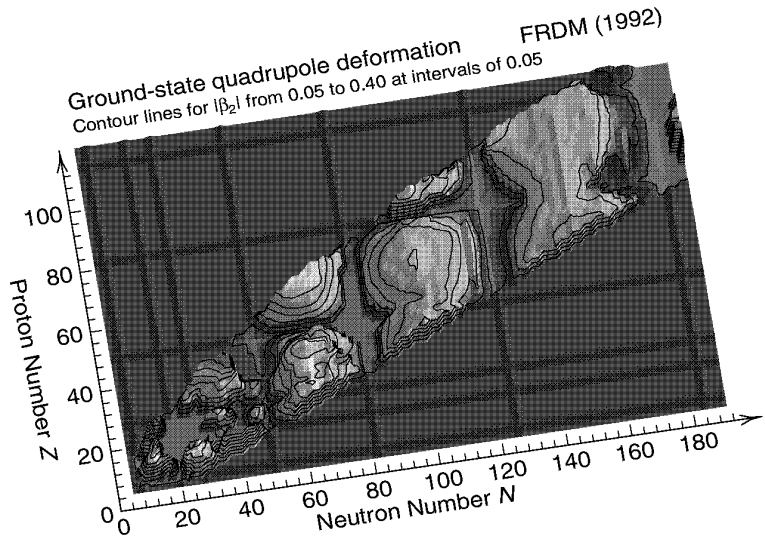
\fig quad fig2 201 {Calculated quadrupole deformations of
even-even nuclei, illustrating the transitions from spherical to deformed nuclei
as one moves away from magic numbers.}

\section{Ground-State Deformations}\label{def}

In our calculations, we specify a general nuclear shape in terms of deviations
from a spheroidal shape by use of Nilsson's
$\epsilon$~parameterization.$\,$\cite{Ni}\ess The ground-state shape is
determined by initially minimizing the nuclear potential energy of deformation
with respect to the two symmetric shape coordinates $\epsilon_2$ and
$\epsilon_4$. During this minimization, we include a prescribed smooth
dependence of the higher symmetric deformation $\epsilon_6$ on the two
independent coordinates $\epsilon_2$ and $\epsilon_4$.  This dependence is
determined by minimizing the macroscopic potential energy of $^{240}$Pu with
respect to $\epsilon_6$ for fixed values of $\epsilon_2$ and $\epsilon_4$.  We
then vary separately $\epsilon_6$ and the mass-asymmetric, or octupole,
deformation $\epsilon_3$, with $\epsilon_2$ and $\epsilon_4$ held fixed at their
previously determined values, to calculate any additional lowering in energy
from these two degrees of freedom.

For presentation purposes, it is sometimes more convenient to express the
nuclear ground-state shape in terms of the $\beta$~parameterization, where the
shape coordinates represent the coefficients in an expansion of the radius
vector to the nuclear surface in a series of spherical harmonics.
Figures~\ref{quad} and \ref{hex} show our calculated quadrupole and hexadecapole
deformations, respectively, in terms of $\beta_2$ and $\beta_4$, which are
determined by transforming our calculated shapes from the
$\epsilon$~parameterization.

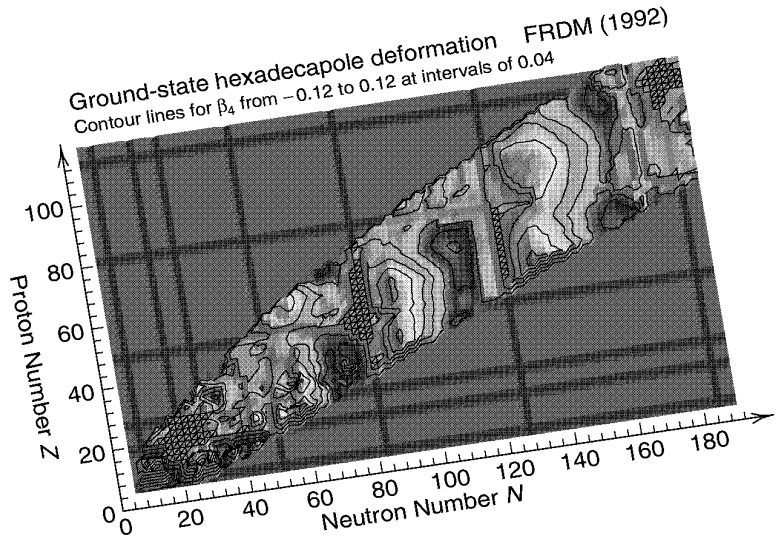
\fig hex fig3 201 {Calculated hexadecapole deformations of
even-even nuclei, illustrating the transitions from bulging to indented
equatorial regions as one moves from smaller to larger magic numbers.}

The inclusion of the $\epsilon_6$ and $\epsilon_3$ shape degrees of freedom is
crucial for the isolation of such physical effects as the Coulomb redistribution
energy, which arises from a central density depression.$\,$\cite{MNMS2}\ess As
illustrated in Fig.~\ref{eps6}, an independent variation of the symmetric
deformation $\epsilon_6$ is important for several regions of nuclei.  For
even-even nuclei, the maximum reduction in energy relative to that for a
prescribed smooth $\epsilon_6$ dependence is 1.28~M\eV and occurs for
$^{252}$Fm.  As illustrated in Fig.~\ref{eps3}, the mass-asymmetric deformation
$\epsilon_3$ is important for nuclei in a few isolated regions.  For even-even
nuclei, the maximum reduction in energy relative to that for a symmetric shape
is 1.29~M\eV and occurs for the neutron-rich nucleus $^{194}$Gd.  For even-even
nuclei close to the valley of $\beta$-stability, the maximum reduction in energy
relative to that for a symmetric shape is 1.20~M\eV and occurs for $^{222}$Ra.

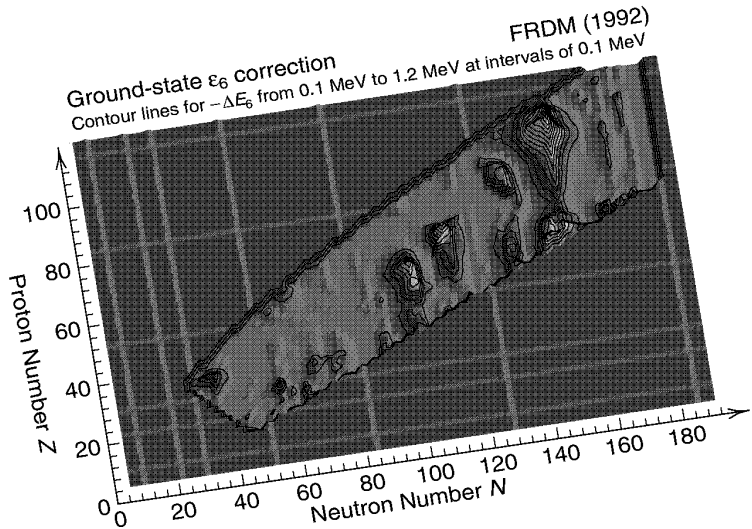
\figt eps6 fig4 198 {Calculated reduction in energy of even-even
nuclei arising from an independent variation in $\epsilon_6$, relative to that
for shapes with a prescribed smooth $\epsilon_6$ dependence.  Note that the sign
of the $\epsilon_6$ correction is reversed in this plot for clarity of display.}
  
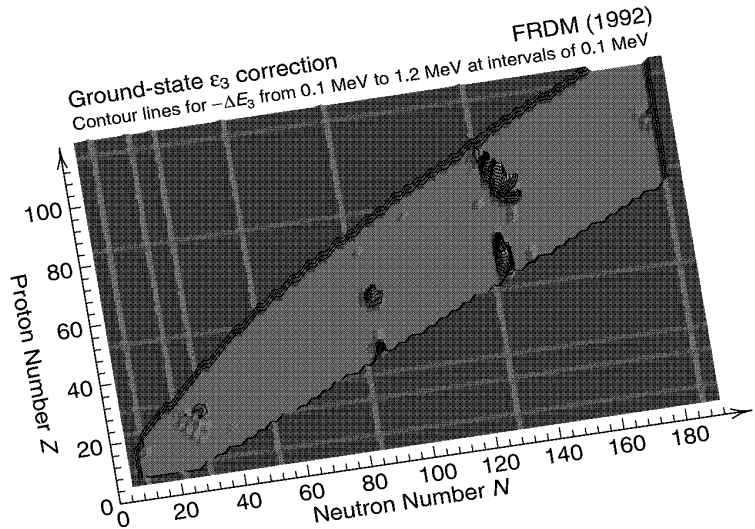
\figt eps3 fig5 198 {Calculated reduction in energy of
even-even nuclei arising from the inclusion of $\epsilon_3$ deformations,
relative to that for symmetric shapes.  Note that the sign of the $\epsilon_3$
correction is reversed in this plot for clarity of display.}

\section{Reliability for Extrapolations to New Regions of Nuclei}\label{extraps}
 
For the original 1654 nuclei included in the adjustment, the theoretical error,
determined by use of the maximum-likelihood method with no contributions from
experimental errors,$\,$\cite{MNMS,MNK}\mss is 0.669 M\eVp Although some large
systematic errors exist for light nuclei, they decrease significantly for
heavier nuclei.

Between 1989 and 1996, the masses of 371 additional nuclei heavier than $^{16}$O
have been measured,$\,$\cite{AW}$^{\sen}\,$\cite{H}\mss which provides an ideal
opportunity to test the ability of mass models to extrapolate to new regions of
nuclei whose masses were not included in the original adjustment.
Figure~\ref{frdm} shows as a function of the number of neutrons from
$\beta$-stability the individual deviations between these newly measured masses
and those predicted by the 1992 finite-range droplet model.  The new nuclei fall
into three categories, with the first category corresponding to 273 nuclei lying
on both sides of the valley of $\beta$-stability.$\,$\cite{AW}\ess The second
category corresponds to 91 proton-rich nuclei produced by fragmentation of
$^{209}$Bi projectiles incident on a thick Be target in the experimental storage
ring (ESR) at the Gesellschaft f\"ur Schwerionenforschung (GSI) in Darmstadt,
Germany.$\,$\cite{K}\ess The third category corresponds to seven proton-rich
superheavy nuclei discovered in the separator for heavy-ion reaction products
(SHIP) at GSI whose masses are estimated by adding the highest $\alpha$-decay
energy release at each step in the decay chain to known masses.$\,$\cite{H}\ess
This procedure could seriously overestimate the experimental masses of some of
the heavier nuclei because different energy releases have been observed in some
cases.$\,$\cite{H}\ess To account for this uncertainty, we have assigned a mass
error of 0.5~M\eV for each of these seven nuclei.  Also, to account for errors
of unknown origin, we have included an additional 0.076~M\eV
contribution$\,$\cite{N} to the mass errors for each of the 91 nuclei in the
second category.  The theoretical error of the 1992 finite-range droplet model
[FRDM (1992)] for all of the 371 newly measured masses is 0.570~M\eVp The
reduction in error arises partly because most of the new nuclei are located in
the heavy region, where the model is more accurate.

Analogous deviations occur for version 1 of the 1992 extended-Thomas-Fermi
Strutinsky-integral [ETFSI-1 (1992)] model of Aboussir, Pearson, Dutta, and
Tondeur.$\,$\cite{APDT}\ess In this model, the macroscopic energy is calculated
for a Skyrme-like nucleon-nucleon interaction by use of an extended Thomas-Fermi
approximation.  The shell correction is calculated from single-particle levels
corresponding to this same interaction by use of a Strutinsky-integral method,
and the pairing correction is calculated for a $\delta$-function pairing
interaction by use of the conventional BCS approximation.  The constants of the
model were determined by adjustments to the ground-state masses of 1492 nuclei
with mass number $A \ge 36$, which excludes the troublesome region from $^{16}$O
to mass number $A = 35$.  The theoretical error corresponding to 1540 nuclei
whose masses were known experimentally$\,$\cite{A} at the time of the original
adjustment is 0.733~M\eVp The theoretical error for 366 newly measured
masses$\,$\cite{AW}$^{\sen}\,$\cite{H} for nuclei with $A \ge 36$ is 0.739~M\eVp

Similar results hold for the 1994 Thomas-Fermi [TF (1994)] model of \linebreak
Myers and Swiatecki.$\,$\cite{MS}\ess In this model, the macroscopic energy is
calculated for a generalized Seyler-Blanchard nucleon-nucleon interaction by use
of the original Thomas-Fermi approximation.  For $N,Z \ge 30$ the shell and
pairing corrections were taken from the 1992 finite-range droplet model, and for
$N,Z \le 29$ a semi-empirical expression was used.  The constants of the model
were determined by adjustments to the ground-state masses of the same 1654
nuclei with $N,Z \ge 8$ ranging from $^{16}$O to $^{263}$106 whose masses were
known experimentally in 1989 that were used in the 1992 finite-range droplet
model.  The theoretical error corresponding to these 1654 nuclei is 0.640~M\eVp
The reduced theoretical error relative to that in the 1992 finite-range droplet
model arises primarily from the use of semi-empirical microscopic corrections in
the extended troublesome region \linebreak $N,Z \le 29$ rather than microscopic
corrections calculated more fundamentally.  The theoretical error for 371 newly
measured masses$\,$\cite{AW}$^{\sen}\,$\cite{H} is 0.620 M\eVp

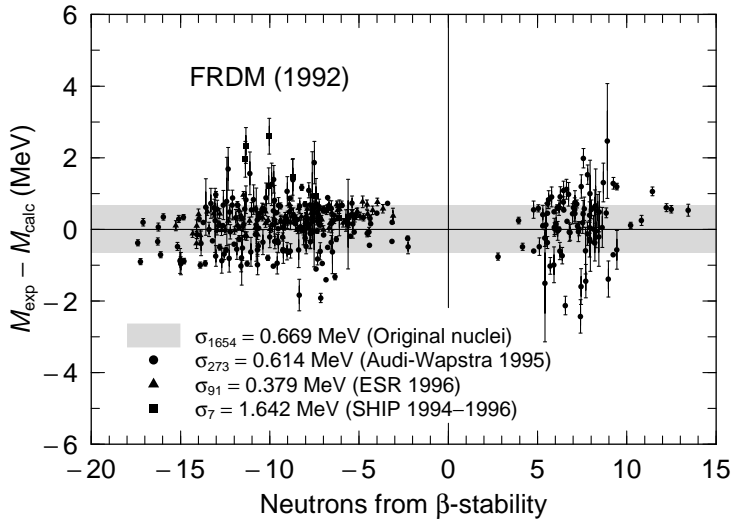
\fig frdm fig6 195 {Deviations between experimental and calculated
masses for 371 new nuclei whose masses were not included in the 1992 adjustment
of the finite-range droplet model.$\,$\cite{MNMS,MNK}\ess}

As summarized in Table~\ref{extrapt}, the theoretical error for the newly
measured masses relative to that for the original masses to which the model
constants were adjusted {\it decreases\/} by 15\% for the FRDM (1992), increases
by 1\% for the ETFSI-1 (1992) model, and {\it decreases\/} by 3\% for the TF
(1994) model.  These macroscopic-microscopic mass models can therefore be
extrapolated to new regions of nuclei with differing amounts of confidence.

\begin{table} 
\caption[extrapt]{Extrapolateability of Three Mass Models to New Regions of
Nuclei.}
\label{extrapt}
\vspace{8pt} 
\begin{center} 
\begin{tabular}{lcccccccc}  
\hline \vspace{-10.150pt} \\
& & \multicolumn{2}{c}{Original nuclei} & & \multicolumn{2}{c}{New nuclei} & & \\[-0.200pt]
\cline{3-4}\cline{6-7}\\[-9.750pt] 

Model & & ${N}_{\rm nuc}$ & Error & & ${N}_{\rm nuc}$ & Error & & Error \\
 
& & & (M\eVpr & & & (M\eVpr & & \hspace{0pt} ratio \vspace{1.275pt} \\
\hline \vspace{-9.675pt} \\

FRDM (1992) & & 1654 & 0.669 & & 371 & 0.570 & & 0.85 \\[6.5pt]

ETFSI-1 (1992) & & 1540 & 0.733 & & 366 & 0.739 & & 1.01 \\[6.5pt]

TF (1994) & & 1654 & 0.640 & & 371 & 0.620 & & \hspace{0pt} 0.97 \vspace{1.275pt} \\

\hline 
\end{tabular} 
\vspace{1pt} 
\end{center}
\end{table}

\section{Rock of Metastable Superheavy Nuclei}\label{rock}

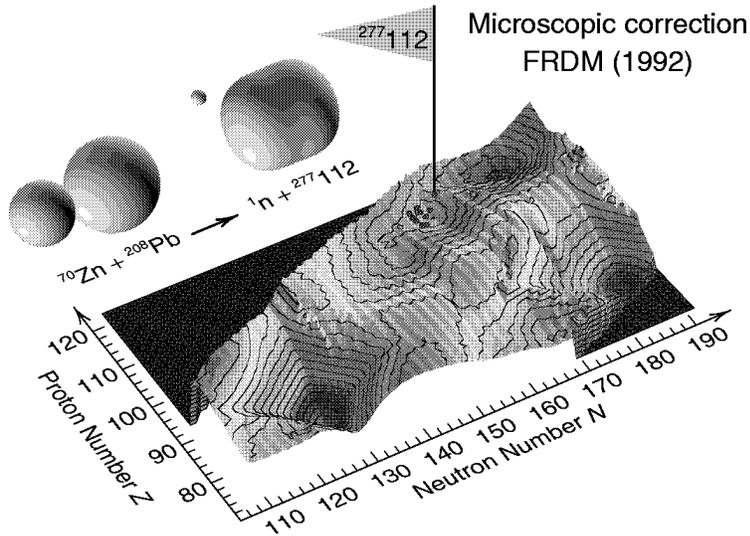
\figt rohic fig7 201 {Ten recently discovered superheavy
nuclei,$\,$\cite{H+}$^{\sen}\,$\cite{O}\mss superimposed on a theoretical
calculation$\,$\cite{MNMS,MNK} of the microscopic corrections to the
ground-state masses of nuclei extending from the vicinity of lead to heavy and
superheavy nuclei.  The heaviest nucleus, whose location on the diagram is
indicated by the flag, was produced through a gentle reaction between spherical
$^{70}$Zn and $^{208}$Pb nuclei in which a single neutron was
emitted.$\,$\cite{H+}\ess}
 
The heaviest nucleus known to man, $^{277}$112, was discovered$\,$\cite{H+} in
February 1996 at the GSI by use of the gentle fusion reaction $^{70}$Zn +
$^{208}$Pb $\rightarrow$~$^1$n~+~$^{277}$112.  It is the latest in a series of
about 10 recently discovered nuclei$\,$\cite{H+}$^{\sen}\,$\cite{O} lying on a
rock of deformed metastable superheavy nuclei predicted to
exist$\,$\cite{MNMS,MNK,MN}$^{\sen}\,$\cite{PS} near the deformed proton magic
number at 110 and deformed neutron magic number at 162.  These 10 superheavy
nuclei are shown in Fig.~\ref{rohic} as tiny deformed three-dimensional objects.
Most of the metastable superheavy nuclei that have been discovered live for only
about a thousandth of a second, after which they generally decay by emitting a
series of alpha particles.  However, the decay products of the most recently
discovered nucleus $^{277}$112 show for the first time that nuclei at the center
of the predicted rock of stability live longer than 10~seconds.

We have used the macroscopic-microscopic method recently to calculate the fusion
barrier for several reactions leading to deformed superheavy
nuclei.$\,$\cite{MNAHM}\ess For the reaction $^{70}$Zn + $^{208}$Pb
$\rightarrow$~$^1$n~+~$^{277}$112, the microscopic shell and pairing corrections
associated primarily with the doubly magic $^{208}$Pb target nucleus lower the
total potential energy at the touching configuration by about 12~M\eV relative
to the macroscopic energy.  These shell and pairing corrections persist from the
touching configuration inward to a position only slightly more deformed than the
ground-state shape.  The resulting maximum in the fusion barrier is about 2~M\eV
lower than the center-of-mass energy that was used in the GSI experiment that
produced $^{277}$112.

One possibility to reach the island of spherical superheavy nuclei near
$^{290}$110 that is predicted to lie beyond our present horizon involves the use
of prolately deformed targets and projectiles that also possess large negative
hexadecapole moments, which leads to large indented equatorial
regions.$\,$\cite{IMNS}\ess

\section{Summary and Conclusion}\label{sum}

The FRDM (1992) and two other macroscopic-microscopic models have been used
recently to calculate the ground-state masses and deformations of nuclei
throughout our known chart and beyond, and the FRDM (1992) has also been used to
simultaneously calculate a wide variety of other nuclear properties.  These
models are useful for extrapolating to new regions of nuclei whose masses were
not included in the original adjustment.  Macroscopic-microscopic models have
also correctly predicted the existence and location of a rock of deformed
metastable superheavy nuclei near $^{272}$110 that has recently been discovered.
Nuclear ground-state masses and deformations will continue to provide an
invaluable testing ground for nuclear many-body theories.  The future challenge
is for fully selfconsistent microscopic theories to predict these quantities
with comparable or greater accuracy.

\section*{Acknowledgments}
This work was supported by the U.~S. Department of Energy.


\section*{References}
\begin{thebibliography}{99}

\bibitem{MNMS} P.~M\"oller, J.~R. Nix, W.~D. Myers, and W.~J. Swiatecki,
\Journal{\ADNDT}{59}{185}{1995}.

\bibitem{MNK} P.~M\"oller, J.~R. Nix, and K.~L. Kratz,
\Journal{\ADNDT}{66}{131}{1997}.

\bibitem{APDT} Y.~Aboussir, J.~M. Pearson, A.~K. Dutta, and F.~Tondeur,
\Journal{\ADNDT}{61}{127}{1995}.

\bibitem{MS} W.~D. Myers and W.~J. Swiatecki, \Journal{\NPA}{601}{141}{1996}.

\bibitem{S} V.~M. Strutinsky, \Journal{\NPA}{122}{1}{1968}.

\bibitem{A} G.~Audi, Midstream Atomic Mass Evaluation, private communication
(1989), with four revisions.

\bibitem{Ni} S.~G. Nilsson, K. Dan.\ Vidensk.\ Selsk.\ Mat.\ Fys. Medd.\ {\bf
29}, 16 (1955).

\bibitem{MNMS2} P.~M\"oller, J.~R. Nix, W.~D. Myers, and W.~J. Swiatecki,
\Journal{\NPA}{536}{61}{1992}.

\bibitem{AW} G.~Audi and A.~H. Wapstra, \Journal{\NPA}{595}{409}{1995}.

\bibitem{K} T.~F. Kerscher, Ph.~D. Thesis, Fakult\"at f\"ur Physik,
Ludwig-Maximilians-Universit\"at M\"unchen (1996).

\bibitem{H} S.~Hofmann, \Journal{\ZPA}{358}{125}{1997}.

\bibitem{N} Yu.~Norikov, private communication (1996).

\bibitem{H+} S.~Hofmann et al., \Journal{\ZPA}{354}{229}{1996}.
  
\bibitem{H+b} S.~Hofmann et al., \Journal{\ZPA}{350}{277}{1995}.
  
\bibitem{H+c} S.~Hofmann et al., \Journal{\ZPA}{350}{281}{1995}.

\bibitem{L+} Yu.~A. Lazarev et al., {\em Phys.\ Rev.\ Lett.\ }{\bf 73}, 624
(1994).

\bibitem{O} Yu.~Ts.\ Oganessian, \Journal{\NPA}{583}{823c}{1995}.

\bibitem{MN} P.~M\"oller and J.~R. Nix, \Journal{\ADNDT}{26}{165}{1981}.

\bibitem{BMNZ} R.~Bengtsson, P.~M\"oller, J.~R. Nix, and Jing-ye Zhang,
Phys.\ Scr.\ {\bf 29}, 402 (1984).

\bibitem{PS} Z.~Patyk and A.~Sobiczewski, \Journal{\PLB}{256}{307}{1991}.

\bibitem{MNAHM} P.~M\"oller, J.~R. Nix, P.~Armbruster, S.~Hofmann, and
G.~M\"unzenberg, \ZPA, in press.

\bibitem{IMNS} A.~Iwamoto, P.~M\"oller, J.~R. Nix, and H.~Sagawa,
\Journal{\NPA}{596}{329}{1996}.

\end{thebibliography}
\end{document}